\documentclass[12pt,a4paper,fleqn,twoside]{article}

\usepackage{amsmath}
\usepackage{amssymb}

\numberwithin{equation}{section} 
\newtheorem{theorem}{Theorem}[section]
\newtheorem{lemma}[theorem]{Lemma}
\newtheorem{definition}[theorem]{Definition}
\newtheorem{example}[theorem]{Example}

\newenvironment{proof}{\paragraph{Proof.}}{\hfill $\square$\\}
\newenvironment{proof*}{\paragraph{Proof.}}{}

\renewcommand{\pmatrix}[2]{\left ( \begin{array}{#1} #2 \end{array} \right )}

\newcommand{\arrow}{\rightarrow}

\newcommand{\bb}[1]{\mathbb{#1}}

\newcommand{\alga}{\mathfrak{a}}
\newcommand{\alg}{\mathfrak{g}}
\newcommand{\Alg}{\mathcal{A}}
\newcommand{\alge}{\mathfrak{e}}

\newcommand{\Cm}{\mathbb{C}}
\newcommand{\smf}{\mathcal{C}^\infty}

\newcommand{\Rm}{\mathbb{R}}

\newcommand{\pr}{\partial}

\newcommand{\me}{\geqslant}

\newcommand{\Dx}[1]{\partial_{x}^{#1}}

\newcommand{\deriv}[2]{\frac{\partial #1}{\partial #2}}
\newcommand{\bra}[1]{\left (#1\right )}
\newcommand{\brac}[1]{\left [#1\right ]}

\newcommand{\pobr}[1]{\left \{#1\right \}}

\newcommand{\Tr}{{\rm Tr}}

\newcommand{\Lg}{\mathcal{L}}
\newcommand{\e}{\mathcal{E}}
\newcommand{\hk}{\hslash}
\newcommand{\Matrix}[3]{\left (
\begin{array}{c}
#2\\ #3
\end{array}
\right )_{#1}}

\newcommand{\Matrixx}[4]{\left (
\begin{array}{c}
#2\\ #3\\ #4
\end{array}
\right )_{#1}}

\newcommand{\var}[2]{\frac{\delta #1}{\delta #2}}
\newcommand{\ad}{{\rm ad}}
\newcommand{\res}{{\rm res}}

\begin{document}

\title{From dispersionless to soliton systems via Weyl-Moyal like deformations.}

\author{Maciej B\l aszak\footnote{E-mail: blaszakm@amu.edu.pl}$\ $ and B\l a\.zej M. Szablikowski\footnote{E-mail: bszablik@amu.edu.pl }\\Institute of Physics, A.Mickiewicz University,\\Umultowska 85, 61-614 Pozna\'n, Poland}



\maketitle

\begin{abstract}
The formalism of quantization deformation is reviewed and the Weyl-Moyal like
deformation is applied to systematic construction of the field and lattice integrable
soliton systems from Poisson algebras of dispersionless systems.\\ 
{\it (To appear in J. Phys. A: Math. Gen.)}

\end{abstract}

\section{Introduction}

Recently, various aspects of the Moyal deformation theory and its application to the
integrable field systems, which leads to the so-called Moyal type Lax dynamics, have
become of increasing interest \cite{Ta}-\cite{TLC}. The aim of this paper is to
present a complete picture of construction of the field and lattice soliton systems by
Weyl-Moyal like deformations from Poisson algebras of underlying dispersionless
systems. The Weyl-Moyal like deformation is the special case of the deformation
quantization.

In the theory of evolutionary systems (dynamical systems) one of the most important
issues is a systematic method for construction of integrable systems. As integrable
systems we understand those which have infinite hierarchy of symmetries and
conservation laws. It is well known that a very powerful tool, called the classical
$R$-matrix formalism, proved to be very fruitful in systematic construction of the
field and lattice soliton systems as well as dispersionless systems (see
\cite{S-T-S}-\cite{BS} and the references there).

The crucial point of the formalism is the observation that integrable dynamical
systems can be obtained from the Lax equations
\begin{equation}\label{laxdyn}
  L_t = \ad_A^* L = \brac{A,L}
\end{equation}
i.e. a coadjoint action of some Lie algebra $\alg$ on its dual $\alg^*$, with the Lax
operators taking values from this Lie algebra $\alg^* \cong \alg$, equipped with the
Lie bracket $[\cdot, \cdot]$. From \eqref{laxdyn} it is clear that we confine to such
algebras $\alg$ for which its dual $\alg^*$, related to the $\alg$ by the duality map
$\langle \cdot,\cdot \rangle \arrow \bb{R}$, can be identified with $\alg$. So, we
assume the existence of a scalar product $\bra{\cdot, \cdot}_\alg$ on $\alg$ which is
symmetric, non-degenerate and $\ad$-invariant:
\begin{equation}
(\ad_a b,c)_\alg + (b,\ad_a c)_\alg = 0.
\end{equation}
This abstract representation \eqref{laxdyn} of integrable systems is referred to as
the Lax dynamics.

On the space of smooth functions on the dual algebra $\alg^*$ there exists a natural
Lie-Poisson bracket
\begin{equation}\label{liepo}
\pobr{H,F}(L):=\langle L,[dF,dH] \rangle \qquad L\in \alg^* \quad H, F\in \smf (\alg^*)
\end{equation}
where $dF$, $dH$ are differentials belonging to $\alg$. A linear map $R:\alg \arrow
\alg$, such that the bracket
\begin{equation}\label{rbra}
  \brac{a,b}_R := \brac{R a, b} + \brac{a,R b}
\end{equation}
is a second Lie product on $\alg$ is called the classical $R$-matrix. A sufficient condition for $R$ to be an $R$-matrix is
\begin{equation}\label{YBa}
  \brac{R a, R b} -R \brac{a, b}_R + \alpha \brac{a,b} = 0\qquad a, b\in \alg
\end{equation}
where $\alpha$ is some real number, called the Yang-Baxter equation YB($\alpha$).

Then, bracket \eqref{rbra} is related to another Lie-Poisson bracket and the
appropriate Poisson tensor as follows
\begin{equation}\label{rliepo}
\pobr{H,F}_1(L):=\langle L,[dF,dH]_R \rangle =: \langle dF, \theta_1 (L) dH \rangle .
\end{equation}
The Casimir functions $C$ of the natural Lie-Poisson bracket \eqref{liepo}, i.e.
\begin{equation}
  \pobr{C,F}=0 \qquad \forall F\in \smf (\alg^*)
\end{equation}
are in involution with respect to the Lie-Poisson bracket \eqref{rliepo}. Hence, the
vector fields generated by such Casimir functions
\begin{equation}\label{evlax}
  L_{t_n} = \theta_1 (L) dC_n = \brac{R \bra{dC_n}, L}
\end{equation}
commute mutually as the map $\theta \circ d$ is a Lie algebra homomorphism.
Moreover \eqref{evlax} are Hamiltonian equations. The hierarchy of evolution equations
\eqref{evlax} is the Lax hierarchy with common infinite set of symmetries and
conserved quantities. In this sense \eqref{evlax} represents a hierarchy of integrable
evolution equations.

It is known that the systems \eqref{evlax} are tri-Hamiltonian with respect to three
Poisson brackets called the linear, quadratic and cubic, reflecting the dependence on
the $L$. The linear tensor $\theta_1 (L)$ takes the form \cite{OR}
\begin{equation}\label{pot1}
\theta_1 (L) dH = -\ad_L R (dH) - R^* \ad_L dH
\end{equation}
where $R^*$ is the adjoint of $R$, i.e. $\bra{R a,b}_\alg = \bra{a, R^* b}_\alg$. The
quadratic case is more complex. A tensor $\theta_2 (L)$ \cite{Su}
\begin{equation}\label{pot2a}
  \theta_2 (L) dH = A_1 (L dH) L - L  A_2 (dH L) + S(dH L)  L - L S^* (L dH)
\end{equation}
defines a Poisson tensor if the linear maps $A_{1,2}: \alg \longrightarrow \alg$ are
skew-symmetric solutions of the YB($\alpha$) \eqref{YBa}, where $\alpha \neq 0$, and
the linear map $S: \alg \longrightarrow \alg$ with adjoint $S^*$ satisfies
\begin{equation}\label{cc}
\begin{split}
&S\bra{\brac{ A_2 a, b} + \brac{a,A_2 b}} = \brac{S a, S b}\\
&S^* \bra{\brac{ A_1 a, b} + \brac{a,A_1 b}} = \brac{S^* a, S^* b}  .
\end{split}
\end{equation}
In the special case when $\frac{1}{2} \bra{R-R^*}$ satisfies the YB($\alpha$), for the
same $\alpha$ as $R$, under the substitution $A_1=A_2 = R-R^*, S=S^*=R+R^*$, the
quadratic Poisson operator \eqref{pot2a} reduces to \cite{OR}
\begin{align}\label{pot2b}
  \theta_2 (L) dH = -\ad_L R \ad^+_L dH-L R^* \ad_L dH - R^*\bra{\ad_L dH} L
\end{align}
where $\ad^+_L A = L A +A L$, and the conditions \eqref{cc} are equivalent to
YB($\alpha$) for $R$. Another special case is when the maps $A_{1,2}$ and $S$
originate from decomposition of $R$-matrix \eqref{rp}
\begin{equation}
  R = A_1 +S = A_2 +S^*
\end{equation}
where $A_{1,2}$ are skew-symmetric. Then the sufficient condition for the Poisson
property of $\theta_2$ is \cite{O}
\begin{equation}
  \brac{A_{1,2} a, A_{1,2} b} + \brac{a,b} = A_{1,2}\bra{\brac{A_{1,2} a,b}+\brac{a,A_{1,2} b}}.
\end{equation}
Finally, the cubic tensor $\theta_3$ takes the form
\begin{equation}\label{pot3}
  \theta_3 (L) dH = -\ad_L R (L dH L)-L R^* (\ad_L dH) L .
\end{equation}

To construct the simplest $R$-structure let us assume that the Lie algebra $\alg$ can be split into a direct sum of Lie subalgebras $\alg_+$ and $\alg_-$, i.e.
\begin{equation}
\alg =\alg_+ \oplus \alg_-\qquad [\alg_\pm,\alg_\pm]\subset \alg_\pm.
\end{equation}
Denoting the projections onto these subalgebras by $P_\pm$, we
define the $R$-matrix as
\begin{equation}\label{rp}
R = \frac{1}{2} (P_+ - P_-).
\end{equation}
It is easy to verify that \eqref{rp} solves YB($\frac{1}{4}$).

Following the above scheme, we are able to construct in a systematic way integrable
Hamiltonian systems, with infinite hierarchy of involutive constants of motion and
infinite hierarchy of related commuting symmetries, once we fix a Lie algebra. For
example the Lie algebra of pseudo-differential operators with the commutator leads to
construction of soliton systems \cite{S-T-S}-\cite{Bl1}. The Lie algebra of shift
operators leads to lattice field systems \cite{K}-\cite{O}. On the other hand, the
Poisson algebras (which are Lie algebras with associative, commutative multiplication)
of formal Laurent series leads to the construction of dispersionless systems
\cite{T-T}-\cite{BS}.

As well known, a quasi-classical limit of field and lattice soliton systems gives
related integrable dispersionless systems. We would like to inverse this procedure and
construct field and lattice soliton systems from some classes of integrable
dispersionless systems through a Weyl-Moyal like deformation quantization procedure.
Actually, we will do it on the level of their Lax representations.

The idea behind the deformation quantization theory \cite{BFFLS}-\cite{Kon} is that an
classical system can be obtained from quantum system by the quasi-classical limit $\hk
\arrow 0$, where $\hk$ is the Planck constant divided by $2 \pi$. Therefore, the
quantization of classical systems should be done by appropriate deformations depending
on a formal parameter $\hk$. The classical fields (observables) belong to the
associative commutative algebra of smooth functions, with standard multiplication,
equipped with the Poisson bracket $\pobr{\cdot , \cdot}$. The idea of deformation
quantization relies on deformation of the usual multiplication to the new associative
but non-commutative product called $\star$-product. It depends on formal parameter
$\hk$, with the assumption that the $\star$-product in the limit $\hk \arrow 0$
reduces to the standard multiplication and also that the Lie bracket
$\pobr{f,g}_\star:=\frac{1}{\hk}(f\star g - g\star f)$, where $f$,$g$ are smooth
functions, reduces to the Poisson bracket. As well known, an arbitrary Poisson tensor,
corresponding to the Poisson bracket, can be written by the wedge product of
appropriate commuting vector fields. Then, the $\star$-product can be easily
constructed by the so-called Weyl-Moyal like deformation. The details will be given in
the next section.

The paper is organized as follows. In Section 1 we have briefly present a number of
basic facts and definitions concerning the classical $R$-matrix formalism. In Section
2 we review the deformation quantization theory and we present the Weyl-Moyal like
deformations. The Poisson algebras of formal Laurent series are introduced in Section
3 and then, in Section 4, the Weyl-Moyal like deformation procedure is applied to
them. In Section 5 we apply the formalism of $R$-matrix to the quantized Poisson
algebras and we illustrate the results with particular examples. Finally, in Section 6
are given some conclusions.

\section{Star products and deformation quantizable Poisson brackets}

Let $\Alg = \mathcal{C}^\infty(M)$ be the space of all smooth ($\Rm$ or $\Cm$ valued) functions on $2n$-dimensional smooth manifold $M$. Let $\pobr{\cdot,\cdot}_{PB}$ be the classical Poisson bracket, which is bilinear, skew-symmetric and satisfies the Jacobi identity. Obviously $\Alg$ is a commutative, associative algebra over $\Rm$ or $\Cm$ with the standard multiplication.

Let $\star$ be the deformed associative non-commutative multiplication on $\Alg$ given by the following formula
\begin{equation}\label{star}
f\star g = \sum_{k\geq 0} \hslash^k B_k(f,g)\qquad f,g\in \Alg
\end{equation}
where $\hk$ is the formal parameter and $B_k:\Alg \times \Alg \longrightarrow \Alg$
are bidifferential (bilinear) operators. We also define deformed bracket as a
commutator
\begin{equation}\label{sbra}
\pobr{f,g}_\star:=\frac{1}{\hk}(f\star g - g\star f).
\end{equation}
\begin{definition}\label{stardef}
An associative deformed multiplication $\star$, given by the formula \eqref{star}, is
a formal quantization of the algebra $\Alg$ and is called the $\star$-product if
\begin{enumerate}\label{starprop}
\item $\lim_{\hk \arrow 0} f\star g = fg$,
\item $c\star f = f\star c = cf$ for $c\in \Rm$ or $\Cm$,
\item $\lim_{\hk \arrow 0} \pobr{f,g}_\star = \pobr{f,g}_{PB}$.  
\end{enumerate}
\end{definition}

\begin{lemma}
The bracket \eqref{sbra} defined by the $\star$-product is bilinear, skew-symmetric
and satisfies the Jacobi identity. So, it is well defined Lie bracket.
\end{lemma}
The proof is obvious as the Jacobi identity is a consequence of an associativity of
multiplication $\star$. Hence, the bracket \eqref{sbra} is called the deformation
quantization of the underlying classical Poisson bracket $\pobr{\cdot,\cdot}_{PB}$.

As follows from definition \eqref{stardef}
\begin{equation}
B_0(f,g)=fg
\end{equation}
and
\begin{equation}\label{B1}
B_1(f,g)-B_1(g,f)=\pobr{f,g}_{PB}.
\end{equation}
The associativity of the $\star$-product implies that the bilinear maps $B_k$ satisfy the
equations
\begin{equation}
\sum_{s=0}^k \left[ B_s(B_{k-s}(f,g),h)-B_s(f,B_{s-k}(g,h))\right]=0\qquad k\geq 1 .
\end{equation}
Hence, $B_1$ satisfies the equation
\begin{equation}
B_1(f,g)h-fB_1(g,h)+B_1(fg,h)-B_1(f,gh) = 0 .
\end{equation}

Let $D:\Alg \longrightarrow \Alg$ be a linear automorphism parametrized by $\hk$, such
that
\begin{equation}\label{auto}
Df = \sum_{k\geq 0} \hk^k D_k f\qquad D_0 = 1
\end{equation}
where $D_k$ are differential operators. Such an automorphism produces a new $\star'$
in $\Alg$ in the following way
\begin{equation}\label{gauge}
f\star' g:= D(D^{-1}f\star D^{-1}g).
\end{equation}
The associativity of the new $\star'$ follows from the associativity of the old
$\star$-product, as
\begin{multline}
f\star'(g\star' h) = f\star' D(D^{-1}g\star D^{-1}h) = D(D^{-1}f\star (D^{-1}g\star D^{-1}h))\\
= D((D^{-1}f\star D^{-1}g)\star D^{-1}h) =  D(D^{-1}f\star D^{-1}g)\star' h = (f\star'
g)\star' h.
\end{multline}
By transformation \eqref{auto} one finds the following expression
\begin{equation}
B_1'(f,g)=B_1(f,g)-fD_1g-D_1g\cdot f+D_1(fg).
\end{equation}
Then
\begin{equation}
B_1'(f,g)-B_1'(g,f)=B_1(f,g)-B_1(g,f)=\pobr{f,g}_{PB}
\end{equation}
and
\begin{equation}
\lim_{\hk \arrow 0} \pobr{f,g}_{\star '} = \lim_{\hk \arrow 0} \pobr{f,g}_\star = \pobr{f,g}_{PB}.
\end{equation}
Hence, the new $\star'$ is the second well defined $\star$-product on $\Alg$.
\begin{definition}
Two $\star$-products: $\star$ and $\star'$ are called gauge equivalent or simply
equivalent if there exist a linear automorphism $D:\Alg \longrightarrow \Alg$
\eqref{auto} such that \eqref{gauge}.
\end{definition}

Let us consider now a Weyl-Moyal like deformations. It is well known that an arbitrary
classical Poisson bracket can be presented in the following form
\begin{equation}\label{class}
\begin{split}
 \pobr{f,g}_{PB} =& f \bra{\sum_{i=1}^{n} Y_i \wedge X_i} g =
  f \bra{\sum_{i=1}^{n}\bra{Y_i\otimes X_i - X_i\otimes Y_i}} g\\ =& \sum_{i=1}^{n} \brac{Y_i(f)X_i(g)-X_i(f)Y_i(g)}
\end{split}
 \end{equation}
where $X_i, Y_i$, i = 1, ..., n are pair-wise commuting vector fields on $2n$
dimensional smooth manifold $M$ and $f, g \in \Alg = \mathcal{C}^\infty(M)$. The
Jacobi identity for \eqref{class} follows from the commutativity of vector fields
$X_i, Y_i$. From relation \eqref{B1} there are two natural deformations of the
classical bracket \eqref{class} induced by
\begin{equation}\label{first}
  B_1 = \frac{1}{2} \sum_{i=1}^{n} Y_i \wedge X_i
\end{equation}
and by
\begin{equation}\label{second}
  B_1' = \sum_{i=1}^{n} Y_i \otimes X_i
\end{equation}
respectively. In what follows, we will use the Einstein summation convention in the case of repeating indices $i$, $j$ at the vectors $X$, $Y$ and a standard convention (with the summation symbols) otherwise. The first case \eqref{first}, leads to the Weyl-Moyal like deformed multiplication
\begin{equation}\label{Moyal}
  f\star g = f \exp \brac{\frac{\hk}{2} Y_i \wedge X_i} g .
\end{equation}
If the classical Poisson bracket \eqref{class} is a canonical one, i.e. $Y_i =
\pr_{p_i}$, $X_i = \pr_{x_i}$ ($\pr_x = \frac{\pr}{\pr x}$, $\pr_p = \frac{\pr}{\pr
p}$), then product \eqref{Moyal} is the Groenewold product \cite{G} and the deformed
bracket \eqref{sbra} is the well known Moyal bracket \cite{M}. Expanding \eqref{Moyal}
one finds
\begin{align}
& f\star g = \sum_{s= 0}^{\infty} \frac{\hk^s}{2^s s!} f  \prod_{k=1}^{k=s} Y_{i_k} \wedge X_{i_k} \ g \notag \\
&= \sum_{s= 0}^{\infty} \frac{\hk^s}{2^s s!} \sum_{m= 0}^s (-1)^m \binom{s}{m}
\bra{Y_{i_1} \ldots Y_{i_{s-m}} X_{j_1} \ldots X_{j_m} f} \bra{Y_{j_1} \ldots Y_{j_m}
X_{i_1} \ldots X_{i_{s-m}} g}.
\end{align}
The second case \eqref{second}, leads to the another Weyl-Moyal like deformed multiplication
\begin{equation}\label{KM}
  f\star g = f \exp \brac{\hk Y_i \otimes X_i} g.
\end{equation}
Again, in the case of the canonical Poisson bracket \eqref{class} the product
\eqref{KM} is the well-known Kupershmidt-Manin (KM) product and the deformed bracket
\eqref{sbra} is the KM bracket \cite{Ma}-\cite{K2}. Expanding \eqref{KM} one finds
\begin{equation}
  f\star g = \sum_{s= 0}^{\infty} \frac{\hk^s}{s!} \bra{Y_{i_1} \ldots Y_{i_s} f} \bra{X_{i_1} \cdots X_{i_s} g}.
\end{equation}

\begin{lemma}
  The product \eqref{KM} is associative. Moreover, it is well defined $\star$-product.
\end{lemma}

Before we prove the lemma, let us introduce the product \eqref{KM} in a little bit different notation
\begin{equation}
  f\star g = \exp \brac{\hk Y_i^f X_i^g} (fg)
\end{equation}
where we use the symbols $Y_i^f$, $X_i^g$ for vector fields acting only on $f$ and $g$, respectively. The following relations for commuting differential operators $X$ and $Y$ are fulfilled:
\begin{align}
&\exp \brac{\hk (X+Y)} = \exp \brac{\hk X} \exp \brac{\hk Y}\\
&\exp \brac{\hk X Y} (f g) = \exp \brac{\hk X (Y^f + Y^g)} (f g)\\
&\exp \brac{\hk X Y} (f g) = f \exp \brac{\hk \bra{XY\otimes 1 + X\otimes Y + Y\otimes X +1\otimes XY}} g .
\end{align}
The first relation is standard and for the second one the proof is as follows
\begin{flalign*}
&\exp \brac{\hk X Y} (f g) = \sum_{s=0}^{\infty} \frac{\hk^s}{s!} X^s Y^s(f g) = \sum_{s=0}^{\infty} \frac{\hk^s}{s!} X^s \sum_{n=0}^{s} \binom{s}{n} \bra{Y^{s-n}f} \bra{Y^n g}\\
&\overset{m=s-n}{=} \sum_{m=0}^{\infty} \frac{\hk^m}{m!} X^m \sum_{n=0}^{\infty}
\frac{\hk^n}{n!} X^n \bra{Y^{m}f} \bra{Y^n g} = \exp \brac{\hk X Y^f } \exp \brac{\hk
X Y^g } (f g).
\end{flalign*}
The last relation follows from the second one as
\begin{flalign*}
 \exp \brac{\hk X Y} (f g) &=   \exp \brac{\hk (X^f+X^g)(Y^f+Y^g)} (f g)\\
&= \exp \brac{\hk \bra{X^fY^f + X^f Y^g + X^g Y^f + X^g Y^g}} (fg) .
\end{flalign*}

\begin{proof} Using the above relations one proves the associativity of \eqref{KM}  product as
\begin{flalign*}
(f\star g) \star h &= \bra{\exp \brac{\hk Y_i^f X_i^g} (fg)} \exp \brac{\hk Y_i \otimes X_i} h\\
&= \exp \brac{\hk Y_i^f X_i^g} \exp \brac{\hk (Y_i^f + Y_i^g) X_i^h } (fgh)\\
&=\exp \brac{\hk Y_i^f (X_i^g + X_i^h) } \exp \brac{\hk Y_i^g X_i^h} (fgh)\\
& = f  \exp \brac{\hk Y_i \otimes X_i} \bra{\exp \brac{\hk Y_i^g X_i^h} (gh)} = f\star (g\star h).
\end{flalign*}
The rest of properties \eqref{starprop} of the $\star$-product is obvious.
\end{proof}

Let us define the linear automorphism $D:\Alg \longrightarrow \Alg$ by
\begin{equation}
  D = \exp \brac{-\frac{\hk}{2} Y_i X_i}\qquad D^{-1} = \exp \brac{ \frac{\hk}{2} Y_i X_i}.
\end{equation}
It relates the $\star$-product \eqref{KM} to the product \eqref{Moyal} by relation
\eqref{gauge} as
\begin{flalign*}
&f\star' g = \exp \brac{-\frac{\hk}{2} Y_i X_i} \bra{ \exp \brac{\frac{\hk}{2} Y_i X_i} (f) \exp \brac{\hk Y_i \otimes X_i} \exp \brac{\frac{\hk}{2} Y_i X_i} (g) }\\
&= f \exp \brac{ -\frac{\hk}{2} \bra{ Y_i X_i \otimes 1 + Y_i\otimes X_i + X_i\otimes Y_i +1\otimes Y_i X_i  }}\\
&\qquad \cdot \exp \brac{\frac{\hk}{2} Y_i X_i\otimes 1 + \hk Y_i \otimes X_i + \frac{\hk}{2} 1\otimes Y_i X_i } g\\
& = f \exp \brac{\frac{\hk}{2} \bra{Y_i \otimes X_i - X_i\otimes Y_i}} g = f \exp \brac{\frac{\hk}{2} Y_i \wedge X_i} g .
\end{flalign*}
Hence, the product \eqref{Moyal} is also a well defined $\star$-product, equivalent to
the $\star$-product \eqref{KM}. Applying to \eqref{KM}
\begin{equation}\label{autoalpha}
  D^\alpha = \exp \brac{- \alpha \frac{\hk}{2} Y_i X_i}
\end{equation}
one finds infinitely many well defined $\star$-products:
\begin{equation}\label{alf}
 f\star^\alpha g = D^\alpha \bra{D^{-\alpha}f\star D^{-\alpha}g} = f \exp \brac{ \frac{\hk}{2} \bra{\bra{2-\alpha} Y_i \otimes X_i -\alpha X_i \otimes Y_i }} g
\end{equation}
where $\alpha \in \mathbb{R}$. All of them are equivalent to etch other and all of
them are quantization of classical Poisson bracket \eqref{class}. Notice, that our
particular $\star$-products \eqref{Moyal} and \eqref{KM} are special cases of
$\star^\alpha$-product \eqref{alf} with $\alpha=1$ and $\alpha=0$, respectively.

Now, we impose the Lie algebra structure on the algebra $\Alg$, denoting it by
$\Alg_\alpha=(\Alg, \star^\alpha)$, with the commutator
\begin{equation}\label{stbralpha}
\pobr{f,g}_{\star^\alpha}:=\frac{1}{\hk}(f\star^\alpha g - g\star^\alpha f).
\end{equation}
Obviously, the automorphism \eqref{autoalpha} induces the isomorphisms between the Lie
algebras
\begin{equation}\label{iso}
  D^{\alpha ' -\alpha}: \Alg_\alpha \longrightarrow \Alg_{\alpha '}
\end{equation}
as
\begin{equation}\label{inv}
  D^{\alpha ' -\alpha} \pobr{f,g}_{\star^\alpha} = \pobr{ D^{\alpha ' -\alpha} f, D^{\alpha ' -\alpha} g}_{\star^{\alpha '}}.
\end{equation}
We will call the Lie algebras $\Alg_\alpha$ gauge equivalent as one can choose freely the algebra one wants to work with.

\section{Poisson algebras of formal Laurent series}

Consider the simplest possible case of $\dim M = 2$, when $M$ is parametrized by a
pair of coordinates $(x, p)$. The Poisson bracket on $\Alg$ can be introduced in
infinitely many ways as
\begin{equation}\label{por}
\pobr{f, g}_{PB}^r := p^r \bra{\deriv{f}{p} \deriv{g}{x} - \deriv{f}{x}
\deriv{g}{p}}\qquad r\in \bb{Z}.
\end{equation}
Moreover, in $\Alg$ there exists the following Poisson subalgebra of formal Laurent
series (Lax polynomials)
\begin{equation}\label{poissonalg}
A =\left \{L = \sum_{i\in \bb{Z}} u_i(x)p^i\right \}
\end{equation}
where the coefficients $u_i$ are smooth functions of $x$. We assume from now on that
$x\in \Omega$, where $\Omega =\bb{S}^1$ if $u_i$ are periodic or $\Omega =\bb{R}$ if
$u_i$ belong to the Schwartz space. An appropriate symmetric product on $A$ is given
by a trace form $(a,b)_A:=\Tr(ab)$:
\begin{equation}\label{trres}
\Tr L = \int_\Omega \res_r L\ dx\qquad \res_r L \equiv u_{r-1}(x)
\end{equation}
which is $\ad$-invariant \cite{Bl2}. In expression \eqref{trres} the integration
denotes the equivalence class of differential expressions modulo total derivatives.
For a given functional $F(L)= \int_\Omega f(u)dx$, we define its differentials as
\begin{equation}\label{res}
dF = \var{F}{L} = \sum_i \var{F}{u_i} p^{r-1-i},
\end{equation}
where $\frac{\delta F}{\delta u}$ is the usual variational derivative.

We construct the simplest $R$-matrix, through a decomposition of $A$ into a direct sum
of Lie subalgebras. For a fixed $r$ let
\begin{equation}
\begin{split}
A_{\me -r+k}&= P_{\me -r+k}A   =\left \{L = \sum_{i\me -r+k} u_i(x)p^i\right \},\\
A_{< -r+k}&= P_{< -r+k}A   =\left \{L = \sum_{i< -r+k} u_i(x)p^i\right \},
\end{split}
\end{equation}
where $P$ are appropriate projections. As presented in \cite{Bl2}, $A_{\me -r+k}, A_{<
-r+k}$ are Lie subalgebras in the following cases:
\begin{enumerate}
\item[1.] $\quad k=0,\ r=0$,
\item[2.] $\quad k=1,2,\ r\in \bb{Z}$,
\end{enumerate}
which one can see through a simple inspection. Then, the $R$-matrix is given by the
projections
\begin{equation}\label{disprmat}
R = \frac{1}{2}(P_{\me -r+k} - P_{< -r+k}) = P_{\me -r+k} - \frac{1}{2} = \frac{1}{2}
- P_{< -r+k}.
\end{equation}
and its adjoint is
\begin{equation}
R^* = \frac{1}{2}(P_{\me -r+k}^* - P_{< -r+k}^*) = \frac{1}{2} - P_{\me 2r-k} = P_{<
2r-k} - \frac{1}{2}.
\end{equation}

Hence, the hierarchy of evolution equations \eqref{evlax} for Casimir functionals
\begin{equation}\label{dispcas}
  C_n(L) = \frac{1}{n+1} Tr \bra{L^{n+1}}\qquad dC_n(L) = L^n
\end{equation}
has the form of two equivalent representations
\begin{equation}\label{displaxh}
L_{t_q} = \pobr{(L^q)_{\me -r+k},L}_{PB}^r = -\pobr{(L^q)_{< -r+k},L}_{PB}^r\qquad
L\in A
\end{equation}
which are Lax hierarchies. Notice that \eqref{displaxh} are multi-Hamiltonian systems
\cite{BS}.

We have to explain what type of Lax operators can be used in (\ref{displaxh}) to
obtain a consistent operator evolution equations equivalent with some nonlinear
integrable dispersionless systems. We are interested in extracting closed systems for
a finite number of fields. To obtain a consistent Lax equation, the Lax operator $L$
has to form a proper submanifold of the full Poisson algebra $A$, i.e. the left and
right-hand sides of expression \eqref{displaxh} have to coincide. They are given in
the form \cite{BS}
\begin{align}
\label{dk0} k=0,\ r=0\ :&\quad L = p^N + u_{N-2} p^{N-2} + ... + u_1 p + u_0\\
\label{dk1} k=1,\ r\in \bb{Z}\ :&\quad L = p^N + u_{N-1} p^{N-1} + ... + u_{1-m}
p^{1-m} +
u_{-m} p^{-m}\\
\label{dk2} k=2,\ r\in \bb{Z}\ :&\quad L = u_N p^N + u_{N-1} p^{N-1} + ... + u_{1-m}
p^{1-m} + p^{-m}
\end{align}
where the $u_i$ are dynamical fields.

\section{Weyl-Moyal like deformation of Poisson algebras of formal Laurent series}

The Poisson brackets \eqref{por} on $\Alg$ can be presented in the following form
\begin{equation}\label{wpor}
\pobr{f, g}_{PB}^r = f \bra{p^r \pr_p \wedge \pr_x} g \qquad r\in \bb{Z}.
\end{equation}
Notice that this is a special case of \eqref{class}, when $Y_1 = p^r \pr_p$ and $X_1 =
\pr_x$ with $\brac{Y_1, X_1}=0$.  For a fixed $r$, the Poisson bracket \eqref{wpor} on
$\Alg$ can be quantized in infinitely many equivalent ways via the
$\star^\alpha$-product \eqref{alf}
\begin{equation}\label{sta}
 f \star^\alpha g = f \exp \brac{ \frac{\hk}{2} \bra{\bra{2-\alpha} p^r\pr_p \otimes \pr_x -\alpha \pr_x \otimes p^r \pr_p }} g .
\end{equation}

One finds that
\begin{equation}
  \bra{p^r\pr_p}^s p^m = c^m_s (r) \  p^{m-s(1-r)}\qquad s \in \mathbb{Z}_+
\end{equation}
where for $k\in \mathbb{Z}$
\begin{equation}
\notag c^{k(1-r)}_s (r) =
   \begin{cases}
    (1-r)^s \frac{k!}{(k-s)!} & \text{for $k\geq s$ and $r\neq 1$}\\
      0                       & \text{for $ s> k \geq 0$ and $r\neq 1$}\\
    (-1+r)^s \frac{(s-k-1)!}{s!}  & \text{for $k < 0$ and $r\neq 1$}\\
   \end{cases}
\end{equation}
for $m \neq k (1-r)$
\begin{equation}
\notag c^m_s (r) = m (m -(1-r))\cdot ... \cdot (m-(s-1)(1-r))
\end{equation}
and for an arbitrary $m\in \mathbb{Z}$
\begin{equation}
\notag c^m_s (1) = m^s .
\end{equation}
One also finds the following relation, which will be useful later
\begin{equation}\label{cms}
  c^m_s (r) = (-1)^s\ c^{(s-1)(1-r)-m}_s (r) .
\end{equation}

Hence, for $\alpha \neq 0, 2$
\begin{equation}
\begin{split}
\notag  &u p^m \star^\alpha v p^n =\\
& \sum_{s=0}^{\infty} \frac{\hk^s}{2^s s!} \sum_{k=0}^s (-1)^k \binom{s}{k} (2-\alpha)^{s-k} \alpha^k \ c^{m}_{s-k}(r) c^{n}_{k}(r) \ u_{kx} v_{(s-k)x} \ p^{m+n-s(1-r)}
\end{split}
\end{equation}
for $\alpha = 0$
\begin{equation}
u p^m \star^0 v p^n = \sum_{s=0}^{\infty} \frac{\hk^s}{s!} c^{m}_{s}(r) \ u v_{sx} \ p^{m+n-s(1-r)}
\end{equation}
and for $\alpha=2$
\begin{equation}
u p^m \star^2 v p^n = \sum_{s=0}^{\infty} \frac{\bra{- \hk}^s}{s!} c^{n}_{s}(r) \
u_{sx} v \ p^{m+n-s(1-r)} .
\end{equation}
Now, a simple inspection leads to the
following relations: for $\alpha \neq 0, 2$
\begin{equation}\label{spo}
    \begin{split}
 & \pobr{u p^m, v p^n}_{\star^\alpha} = \frac{1}{\hk} \bra{u p^m \star^\alpha v p^n - v p^n \star^\alpha u p^m}\\
& \quad = \sum_{s=0}^{\infty} \frac{\hk^{s-1}}{2^s s!} \sum_{k=0}^{s} (-1)^k \binom{s}{k} (2-\alpha)^{s-k} \alpha^k \\
&\qquad \cdot \bra{c^{m}_{s-k}(r) c^{n}_{k}(r) \ u_{kx} v_{(s-k)x} - c^{m}_{k}(r) c^{n}_{s-k}(r) \ u_{(s-k)x} v_{kx}} p^{m+n-s(1-r)} .
   \end{split}
\end{equation}
for $\alpha = 0$
\begin{equation}
  \pobr{u p^m, v p^n}_{\star^0} =
\sum_{s=0}^{\infty} \frac{\hk^{s-1}}{s!} \bra{c^{m}_{s}(r)  u v_{sx} - c^{n}_{s}(r) u_{sx} v} p^{m+n-s(1-r)}
\end{equation}
for $\alpha = 2$
\begin{equation}
  \pobr{u p^m, v p^n}_{\star^2} =
\sum_{s=0}^{\infty} \frac{\hk^{s-1}}{s!} (-1)^s \bra{c^{n}_{s}(r)  u_{sx} v - c^{m}_{s}(r) u v_{sx}} p^{m+n-s(1-r)} .
\end{equation}
So, we can quantize separately the Poisson subalgebra $A$ \eqref{poissonalg} to the
following Lie subalgebras $A_\alpha = (A,\star^\alpha) \subset \Alg_\alpha$.

Obviously, the Lie algebras $A_\alpha$ for a fixed value of $r$ are gauge equivalent
under the isomorphism \eqref{iso}
\begin{equation}\label{lintrans}
  D^{\alpha' - \alpha} : A_\alpha \longrightarrow A_{\alpha'}\qquad D^{\alpha' - \alpha} = \exp \brac{(\alpha -\alpha')\frac{\hk}{2} p^r \pr_p \pr_x}.
\end{equation}
Let
\begin{equation}
 L = \sum_{m= - \infty}^{+\infty} u_m p^m \in A_\alpha \qquad L'=\sum_{n= - \infty}^{+\infty} v_n p^n \in A_{\alpha'}
\end{equation}
then $L' = D^{\alpha' - \alpha} L$ and fields $u_m$, $v_n$ are mutually related in the following way
\begin{equation}\label{mutrel}
  v_n =\sum_{s\geq 0} \bra{(\alpha -\alpha')\frac{\hk}{2}}^s \frac{1}{s!}\ c_s^{s(1-r)+n}(r)\ (u_{s(1-r)+n})_{sx} .
\end{equation}
Because of the gauge equivalence between the Lie algebras $A_\alpha$ we can choose one
Lie algebra with a fixed value of $\alpha$, make all necessary calculations, and then
reconstruct all results for $A_{\alpha '}$ directly from the transformation
\eqref{mutrel}.

On the other hand one can show the following relations
 \begin{align}
\label{rule0}  &u\star^\alpha v = u v\\
  &p^m \star^\alpha p^n = p^{m+n}\\
\label{rule1}  &p^m \star^\alpha u = \sum_{s\geq 0} \frac{\hk^s}{s!}   u_{sx}\star^\alpha \bra{p^r \pr_p}^s p^m\\
\label{rule2}  &u\star^\alpha p^m = \sum_{s\geq 0} \frac{\hk^s}{s!} \bra{p^r \pr_p}^s p^m \star^\alpha u_{sx}.
 \end{align}
As all relations \eqref{rule0}-\eqref{rule2} have the same form independently of
$\alpha$ we skip this index in further considerations. Hence, we can quantize
separately the algebra $A$ to the following special algebra of Lax operators:
\begin{equation}\label{alg}
\alga = \pobr{L = \sum_{i\in \bb{Z}} u_i(x)\star p^i} .
\end{equation}
It is obviously associative algebra under commutation rules
\eqref{rule1}-\eqref{rule2}. The algebra $\alga$ in the case of $r=0$ was considered
for the first time in \cite{DP}. Then, the Lie-bracket on $\alga$ is given by
\begin{equation}\label{starpo}
    \begin{split}
  \pobr{u \star p^m, v \star p^n}_\star &= \frac{1}{\hk} \bra{u\star p^m \star v\star p^n - v\star p^n \star u\star p^m}\\
 &= \sum_{s=0}^{\infty} \frac{\hk^{s-1}}{s!} \brac{c^{m}_{s}(r)\ u v_{sx} - c^{n}_{s}(r)\  u_{sx} v}\star p^{m+n-s(1-r)} .
    \end{split}
\end{equation}
Notice that the algebra $\alga$ differs from that defined in the third section, where we introduced deformation quantization, as in \eqref{alg} we also deformed the Lax polynomials.
Let us remark that the algebra $\alga$ is naturally isomorphic to the algebra $A_0$ as
$u\star^0 p^m = u p^m$. Hence, in the further considerations we will concentrate only
on the algebra $\alga$, as the results for the algebras $A_\alpha$ for all values of
$\alpha$ can be obtained simply by transformations \eqref{mutrel} from $A_0$. The
second reason is that $\alga$ can be considered as the generalization of the algebra
of the pseudo-differential operators and the algebra of the shift operators in the
following sense.

Let us consider the case of $r=0$, then the rules \eqref{rule1} and \eqref{rule2} take the particular form
\begin{align}
&p^m\star u = \sum_{s=0} \hk^s \binom{m}{s} u_{sx}\star p^{m-s},\\
&u\star p^m = \sum_{s=0} (-\hk)^{s} \binom{m}{s} p^{m-s}\star u_{sx}.
\end{align}
and the Lie bracket \eqref{starpo} is
\begin{equation}\label{starpor0}
    \pobr{u \star p^m, v \star p^n} = \sum_{s=0}^{\infty} \hk^{s-1} \brac{\binom{m}{s} u v_{sx} - \binom{n}{s} u_{sx} v}\star p^{m+n-s}.
  \end{equation}
Hence, the algebra $\alga$ for fixed $r=0$ is isomorphic to the algebra of pseudo-differential operators:
\begin{equation}\label{pdo}
\alg = \pobr{\Lg = \sum_{i\in \bb{Z}} u_i(x)\pr_x^i}
\end{equation}
where the multiplication of two such operators uses the generalized Leibniz rule
\begin{equation}
\pr^m u = \sum_{s=0} \hk^s \binom{m}{s} u_{sx} \pr_x^{m-s}\qquad u \pr^m = \sum_{s=0} \bra{-\hk}^s \binom{m}{s} \pr_x^{m-s} u_{sx}
\end{equation}
where $\hk$ is a formal parameter. The Lie algebra structure of $\alg$ is given by the
bracket $\brac{\Lg_1, \Lg_2}=\frac{1}{\hk}(\Lg_1 \Lg_2 -\Lg_2 \Lg_1)$. The isomorphism
is given by the function $sym:\alg \arrow \alga$
\begin{equation}
sym (\Lg) = sym \bra{\sum_{i} u_i(x)\pr_x^i} = \sum_{i} u_i(x)\star p^i = L .
\end{equation}
It has the important property that for arbitrary $\Lg_1, \Lg_2 \in \alg$
\begin{equation}\label{s1}
sym (\Lg_1 \Lg_2 ) = sym(\Lg_1)\star sym(\Lg_2 ) .
\end{equation}
Then it follows that
\begin{equation}\label{s2}
sym (\brac{\Lg_1, \Lg_2} ) = \pobr{sym(\Lg_1), sym(\Lg_2) }_\star = \pobr{L_1, L_2}_\star .
\end{equation}
Hence, $sym$ is the Lie algebra isomorphism. Obviously such Lie algebras $\alg$ for all values of $\hk$ are in natural way isomorphic to the standard algebra of pseudo-differentials operators ($\hk=1$).

Let us now consider the case of $r=1$, then the rules \eqref{rule1} and \eqref{rule2}
become
\begin{align}
\notag p^m\star u(x) &= \sum_{s=0} \frac{\hk^s}{s!} m^s \bra{u(x)}_{sx}\star p^{m}\\
 &=: \e^{m } u(x)\star p^m = u(x+m\hk)\star p^m\\
\notag u(x)\star p^m &= \sum_{s=0} \frac{(-\hk)^{s}}{s!} m^s p^{m}\star \bra{u(x)}_{sx}\\
&=: p^m\star \e^{-m} u(x) = p^m \star u(x-m \hk)
\end{align}
where we use the formula for Taylor expansion and we consider $\e$ as the shift
operator. The Lie bracket \eqref{starpo} is
\begin{equation}\label{starpor1}
    \pobr{u(x) \star p^m, v(x) \star p^n}_\star = \frac{1}{\hk} \brac{ u(x)  v(x+m\hk) - u(x+n\hk) v(x)}\star p^{m+n}.
  \end{equation}
Hence, the algebra $\alga$ for a fixed $r=1$ is isomorphic to the algebra of shift
operators:
\begin{equation}\label{so}
\alge = \pobr{\Lg = \sum_{i\in \bb{Z}} u_i(x) E^i}
\end{equation}
where $E$ is the shift operator such that
\begin{equation}
E^m u(x) = u(x+m\hk) E^m\qquad u(x) E^m = E^m u(x-m\hk)
\end{equation}
where $\hk$ is a formal parameter. The Lie algebra structure of $\alge$ is given by the bracket $\brac{\Lg_1, \Lg_2}=\frac{1}{\hk}(\Lg_1 \Lg_2 -\Lg_2 \Lg_1)$. The isomorphism is given by the function $sym:\alge \arrow \alga$
\begin{equation}
sym (\Lg) = sym \bra{\sum_{i} u_i(x) E^i} = \sum_{i} u_i(x)\star p^i = L .
\end{equation}
Like in the previous case, the relations \eqref{s1} and \eqref{s2} are fulfilled for
arbitrary $\Lg_1, \Lg_2 \in \alge$.

Let us investigate for a moment some properties of the Lie algebra $\alga$. The first observation is the existence of a symmetric, non-degenerate and $\ad$-invariant product on $\alga$ allowing us to identify $\alga$ with its dual $\alga^*$.
\begin{lemma}
An appropriate scalar product on $\alga$ is given by a trace form
\begin{equation}\label{symprod}
  \bra{L_1, L_2}_\alga := \Tr \bra{L_1\star L_2}
\end{equation}
where
\begin{equation}\label{trace}
\Tr L = \int_\Omega \res_r L\ dx,\qquad \res_r L \equiv u_{r-1}(x) .
\end{equation}
Then \eqref{symprod} is symmetric, non-degenerate and $\ad$-invariant.
\end{lemma}
\begin{proof*}
The non-degeneracy of the product \eqref{symprod} is obvious. Let $L_1=\sum_{m} u_{m}\star p^{m}$, $L_2 =\sum_{n} v_{n}\star p^{n}$, then using relations \eqref{rule1} and \eqref{cms} we find
\begin{align*}
  \bra{L_1, L_2}_\alga &= \Tr \bra{\sum_{m, n} u_{m}\star p^{m}\star v_{n}\star p^{n}}\\
&= \Tr \bra{\sum_{m, n} \sum_{s=0}^{\infty} \frac{\hk^s}{s!} c^{m}_{s}(r)\ u_m \bra{v_n}_{sx} \star p^{m+n-s(1-r)}}\\
&= \int_\Omega \sum_n \sum_{s=0}^{\infty} \frac{\hk^s}{s!} c^{(s-1)(1-r)-n}_{s}(r)\ u_{(s-1)(1-r)-n} \bra{v_n}_{sx}\ dx\\
&= \int_\Omega \sum_n \sum_{s=0}^{\infty} \frac{\hk^s}{s!} (-1)^s c^{(s-1)(1-r)-n}_{s}(r)\ \bra{u_{(s-1)(1-r)-n}}_{sx} v_n\ dx\\
&= \int_\Omega \sum_n \sum_{s=0}^{\infty} \frac{\hk^s}{s!} c^n_s(r)\ \bra{u_{(s-1)(1-r)-n}}_{sx} v_n\ dx\\
&= \Tr \bra{\sum_{m, n} \sum_{s=0}^{\infty} \frac{\hk^s}{s!} c^{n}_{s}(r)\ \bra{u_m}_{sx} v_n \star p^{m+n-s(1-r)}}\\
&= \Tr \bra{\sum_{m, n} v_{n}\star p^{n}\star u_{m}\star p^{m}} = \bra{L_2, L_1}_\alga
\end{align*}
where we have used the integration by parts. The $\ad$-invariance follows from
associativity of $\star$-product and symmetry of the product \eqref{symprod} as
\begin{align*}
  \bra{\pobr{A,B}_\star, C}_\alga &= Tr \bra{\frac{1}{\hk} \bra{A\star B\star C -B\star A\star C}}\\
   &= Tr \bra{\frac{1}{\hk} \bra{B\star C\star A -C\star B\star A}} = \bra{\pobr{B,C}_\star, A}_\alga .\tag*{$\square$}
\end{align*}
\end{proof*}

As a consequence, for operators $L = \sum_{i} u_i\star p^i$, the vector fields
$\frac{d}{dt}L \equiv L_t$ and differentials $dH(L)$ are conveniently parameterized by
\begin{equation}\label{grad}
  L_t = \sum_i (u_i)_t \star p^i\qquad dH(L) = \var{H}{L} = \sum_i p^{r-1-i}\star \var{H}{u_i}
\end{equation}
where $\var{H}{u_i}$ is the usual variational derivative of a functional $H =
\int_\Omega h(u,u_x,...)\ dx$. In these frames the trace duality assumes the usual
Euclidean form
\begin{equation}
  \bra{dH, L_t}_\alga = \Tr \bra{dH\star L_t} = \sum_i \int_\Omega \var{H}{u_i} (u_i)_t \ dx .
\end{equation}

Now, one can simply rewrite the trace formula from $\alga$ to
$A_0$ as $\alga\cong A_0$. Then, appropriate trace formulas on
$A_\alpha$ for Lax polynomials $L = \sum_{n} u_n p^n$ are given by
\begin{equation}\label{t1}
\Tr L = \int_\Omega \res_r L\ dx,\qquad \res_r L \equiv u_{r-1}(x)
\end{equation}
which are well defined, as the trace formula is invariant under transformations
\eqref{lintrans} since from \eqref{mutrel} it follows that $v_{r-1}= u_{r-1}$. Hence,
the scalar products take the form
\begin{equation}\label{t2}
  \bra{L_1, L_2}_{A_\alpha} := \Tr \bra{L_1\star^\alpha L_2}.
\end{equation}
and differentials $dH(L)$ are conveniently parameterized by
\begin{equation}\label{t3}
dH(L) = \var{H}{L} = \sum_n p^{r-1-n}\star^{\alpha} \var{H}{u_n}.
\end{equation}
Notice that in formulas \eqref{t1}-\eqref{t3} one has to use the explicit form of
$\star^\alpha$-products.

\section{R-matrix formalism and Lax hierarchies for Lie algebra $\alga$}

To construct the integrable field systems one has to split the algebra $\alga$ into a
direct sum of Lie subalgebras. Observing \eqref{starpo} one finds that in general it
can be done only for $r=0$ or $r=1$. Let us remark that it is possible to choose a Lie
subalgebra of $\alga$ in the form
\begin{equation}
  \pobr{L = \sum_{i\in \bb{Z}} u_i(x)\star p^{i(1-r)}} \qquad r\neq 1
\end{equation}
which can be further split into a direct sum of Lie subalgebras, but this case is
simply related by the transformation $p' = p^{1-r}, x'=\frac{1}{1-r} x$ to the algebra
$\alga$ for the case of $r=0$.

Now, we decompose $\alga$ for $r=0, 1$ into a direct sum of Lie subalgebras in the following way. Let
\begin{equation}
\begin{split}
\alga_{\me -r+k}&= P_{\me -r+k}\alga   =\left \{L = \sum_{i\me -r+k} u_i(x)\star p^i\right \}\\
\alga_{< -r+k}&= P_{< -r+k}\alga   =\left \{L = \sum_{i<-r+k} u_i(x)\star p^i\right \}
\end{split}
\end{equation}
where $P$ are appropriate projections. Then, $\alga_{\me -r+k}, \alga_{< -r+k}$ are
Lie subalgebras in the case of $r=0$ for $k=0,1,2$ and in the case of $r=1$ for $k=1,
2$. Hence, the $R$-matrix is given by the projections
\begin{equation}\label{rmat}
R = \frac{1}{2}(P_{\me -r+k} - P_{< -r+k}) = P_{\me -r+k} -
\frac{1}{2} = \frac{1}{2} - P_{< -r+k}.
\end{equation}
and its adjoint is
\begin{equation}
R^* = \frac{1}{2}(P_{\me -r+k}^* - P_{< -r+k}^*) = \frac{1}{2} -
P_{\me 2r-k} = P_{< 2r-k} - \frac{1}{2}.
\end{equation}

The hierarchy of evolution equations are generated by the Casimir functionals
\begin{equation}\label{cas}
  C_n(L) = \frac{1}{n+1} Tr \bra{L^{n+1}}\qquad dC_n(L) = L^n\qquad L^n = L\star L\star ...\star L
\end{equation}
and for appropriate $k$ has the form of two equivalent representations
\begin{equation}\label{laxh}
L_{t_n} = \pobr{(L^n)_{\me -r+k},L}_\star = -\pobr{(L^q)_{<-r+k},L}_\star
\end{equation}
which are Lax hierarchies.

The Lie algebras $A_\alpha$ can be decompose into direct sum of Lie subalgebras exactly in the same way as $A_0\cong \alga$. Hence, the $R$-matrix \eqref{rmat} is invariant under transformations \eqref{lintrans}. Moreover, as transformations \eqref{lintrans} are Lie algebra isomorphisms \eqref{inv} the Lax hierarchies \eqref{laxh} are also invariant with respect to them.

To construct (1+1)-dimensional closed systems with a finite number of fields we have
to choose properly restricted Lax operators $L$ which give consistent Lax equations
\eqref{laxh}. To obtain a consistent Lax equation, the Lax operator $L$ has to form a
proper submanifold of the full Poisson algebra under consideration, i.e. the left and
right-hand sides of expression \eqref{laxh} have to lie inside this submanifold. In
the case of $r=0$ the admissible simplest restricted Lax operators are given in the form
\begin{align}
\label{laxk0}
k=0: &\quad L = p^N + u_{N-2}\star p^{N-2} + ... + u_1 \star p + u_0\\
\label{laxk1} k=1: &\quad L = p^N + u_{N-1} \star p^{N-1} + ... + u_0 + p^{-1}\star u_{-1}\\
\label{laxk2} k=2: &\quad L = u_N \star p^N + u_{N-1} \star p^{N-1} + ... + u_0 + p^{-1} \star u_{-1} + p^{-2} \star u_{-2} .
\end{align}
In the case of $r=1$ the admissible simplest restricted Lax operators are
\begin{align}
\label{slaxk1} k=1:&\quad L = p^N + u_{N-1}\star p^{N-1} + ... + u_{1-m}\star p^{1-m} + u_{-m}\star p^{-m}\\
\label{slaxk2} k=2:&\quad L = u_N\star p^N + u_{N-1}\star p^{N-1} + ... + u_{1-m}\star p^{1-m} + p^{-m}.
\end{align}

We will now compare the Lax operators related to soliton systems with
the Lax operators related to the dispersionless systems. As follows, the class of
operators \eqref{laxk0} is the same as the class of dispersionless operators
\eqref{dk0}. Hence, all dispersionless systems for $r=0$ and $k=0$ have counterpart
soliton systems. For $r=0$ and $k=1, 2$ the classes of dispersionless Lax operators
are wider. The operators \eqref{laxk1} by the quasi-classical limit ($\hk \rightarrow
0$) reduce to the operators \eqref{dk1} for $m=1$. The operators \eqref{laxk2} reduce
to \eqref{dk2} for $m=2$ but the field $u_{-2}$ by the quasi-classical limit becomes
time independent. For $r=1$ the classes of operators \eqref{slaxk1}, \eqref{slaxk2}
and \eqref{dk1}, \eqref{dk2} are the same, respectively. Thus, all of them have the
counterpart lattice field systems. The remaining dispersionless systems for $r\neq
0,1$ and some for $r=0$ do not have counterpart soliton systems in the quantization
scheme considered.

The evolution systems \eqref{laxh}, with the Casimir
functionals \eqref{cas} as Hamiltonian functions, are tri-Hamiltonian
\begin{equation}\label{laxhtri}
 L_{t_n} = \theta_1(L)dC_n = \frac{1}{2} \theta_2(L)dC_{n-1} =  \theta_3(L)dC_{n-2}
\end{equation}
as it was for the algebra of pseudo-differential operators and the algebra of shift
operators. Nevertheless, as we work with restrictions \eqref{laxk0}-\eqref{slaxk2}, a reduction procedure for the Hamiltonian structures of the general representations \eqref{laxhtri} will be necessary.

\paragraph{The case of $r=0$.}

All Lax operators \eqref{laxk0}-\eqref{laxk2} form a proper submanifold with respect to the linear Poisson
tensor \eqref{pot1} which is given for $R$-matrix \eqref{rmat} in two equivalent
representations
\begin{align}
\notag \theta_1(L)dH &= \pobr{(dH)_{\me k},L}_\star- (\pobr{dH,L}_\star)_{\me -k}\\
&= -\pobr{(dH)_{< k},L}_\star+(\pobr{dH,L}_\star)_{< -k}\qquad k=0,1,2 .
\end{align}
Since $(\var{H}{L})_{\me k}=0$ for $k=0,1$, the linear Poisson tensor for these cases is given in simpler form
\begin{equation}\label{p1r0k}
\theta_1 \bra{\var{H}{L}} = \bra{\pobr{L,\var{H}{L}}_\star}_{\me -k}.
\end{equation}

The quadratic bracket for $k=0$ is given by \eqref{pot2b}
\begin{align}\notag
\theta_2(L)dH &= \hk \pobr{L,\bra{\pobr{dH,L}^+_\star}_{\me 0}}_\star- \hk\pobr{L,\bra{\pobr{dH,L}_\star}_{\me 0}}^+_\star\\
\label{pb1} &= -\hk \pobr{L,\bra{\pobr{dH,L}^+_\star}_{< 0}}_\star+
\hk\pobr{L,\bra{\pobr{dH,L}_\star}_{< 0}}^+_\star
\end{align}
and can be properly restricted to a subspace of the form
\begin{equation}
L = p^N + u_{N-1}\star p^{N-1} + u_{N-2}\star p^{N-2} + ... + u_1 \star p + u_0
\end{equation}
hence a Dirac reduction $u_{N-1}=0$ is required with the final result
\begin{equation}\label{p2r0k0}
\theta_2^{red}\bra{\var{H}{L}} = \frac{1}{\hk}\brac{ \bra{L\star \var{H}{L}}_{\me
0}\star L - L\star \bra{\var{H}{L}\star L}_{\me 0}} +
\frac{\hk}{N}\pobr{\partial_x^{-1}\res \pobr{\var{H}{L},L}_\star,L}_\star
\end{equation}
where $\theta_2^{red}(L)$ is compatible with the linear one \eqref{p1r0k}. For $k=1$,
the quadratic tensor $\theta_2 (L)$ is given by \eqref{pot2a}
\begin{equation}\label{pb2a}
  \theta_2 (L) dH = \frac{1}{\hk} \brac{A_1 (L\star dH)\star L - L\star  A_2 (dH\star L)
   + S(dH\star L)\star  L - L\star S^* (L\star  dH)}
\end{equation}
where
\begin{align}
\notag &A_1(b) = b_{\me 1} -b_0+b_{-1}-b_{< -1}-2\pr_x^{-1} \res \ b\qquad b\in \alga\\
&A_2(b) = b_{\me 0} - b_{< 0}+2 \pr_x^{-1} \res \ b\\
\notag &S(b) = -2 b_{-1} + 2\pr_x^{-1} \res \ b\qquad S^* (b) = -2 b_{0} - 2\pr_x^{-1}
\res \ b
\end{align}
satisfy \eqref{cc}. The Poisson tensor \eqref{pb2a} admits a proper restriction to Lax operators of the form \eqref{laxk1}. Hence, we have
\begin{align}\notag
\theta_2\bra{\var{H}{L}} =& \frac{1}{\hk}\brac{ \bra{L\star \var{H}{L}}_{\me
1}\star L - L\star \bra{\var{H}{L}\star L}_{\me 0} + L\star \bra{L\star
\var{H}{L}}_0}\\ \label{p2r0k1}
 &- \pr_x^{-1} \res \! \bra{\pobr{\var{H}{L},L}_\star}\star L +\hk
\pobr{\pr_x^{-1} \res \pobr{\var{H}{L},L}_\star,L}_\star .
\end{align}
For $k=2$, contrary to the previous cases, we still do not know the proper form of the
quadratic tensor $\theta_2$.

The restricted Lax operators \eqref{laxk0}-\eqref{laxk2} do not form proper submanifolds
with respect to the cubic Poisson tensor \eqref{pot3}
\begin{align}
\notag \theta_3(L)dH &= \pobr{(L\star dH\star L)_{\me k},L}_\star- L\star (\pobr{dH,L}_\star)_{\me -k}\star L\\
&= -\pobr{(L\star dH\star L)_{< k},L}_\star+ L\star (\pobr{dH,L}_\star)_{< k}\star
L\qquad k=0,1,2 .
\end{align}
Nevertheless, the Dirac reduction can be applied. Here, contrary to the previous
cases, the number of constraints depends on $N$, so the reduction has to be considered
separately for each $N$.

\paragraph{The case of $r=1$.}

Both Lax operators \eqref{slaxk1}-\eqref{slaxk2} form a proper submanifold with respect to the linear Poisson
tensor \eqref{pot1}
\begin{align}
\notag \theta_1(L)dH &= \pobr{(dH)_{\me -1+k},L}_\star- (\pobr{dH,L}_\star)_{\me 2-k}\\
\label{p1r1k} &= -\pobr{(dH)_{< -1+k},L}_\star+(\pobr{dH,L}_\star)_{< 2-k}\qquad k=1,2
.
\end{align}
Hence, no additional restrictions are needed.

The quadratic tensor is given by the special case \eqref{pot2b} nevertheless the Lax
operators \eqref{slaxk1}-\eqref{slaxk2} do not form a proper submanifolds. Actually, the
proper submanifold is
\begin{equation}
L = u_N\star p^N + u_{N-1}\star p^{N-1} + ... + u_{1-m}\star p^{1-m} + u_{-m}\star
p^{-m} .
\end{equation}
Thus for $k=1$ the Dirac constraint $u_N=1$ gives
\begin{align}\notag
\theta_2^{red}\bra{\var{H}{L}} =& \hk \pobr{\bra{\pobr{L,\var{H}{L}}_\star^+}_{\me
0},L}_\star-\hk
\pobr{L,\bra{\pobr{\var{H}{L},L}_\star}_{\me 1}}_\star^+\\
\label{p2r1k1} & +\hk \pobr{ \bra{1+\e^{-N}} \bra{1-\e^{-N}}^{-1} \res
\pobr{\var{H}{L},L}_\star,L}_\star
\end{align}
and for $k=2$ with Dirac constraint $u_{-m}=1$ we get
\begin{align}\notag
\theta_2^{red}\bra{\var{H}{L}} =& \hk
\pobr{\bra{\pobr{L,\var{H}{L}}_\star^+}_{\me 1},L}_\star-\hk \pobr{L,\bra{\pobr{\var{H}{L},L}_\star}_{\me 0}}_\star^+ \\
\label{p2r1k2} & -\hk \pobr{ \bra{1+\e^{m}} \bra{1-\e^{m}}^{-1} \res
\pobr{\var{H}{L},L}_\star,L}_\star.
\end{align}

The restricted Lax operators \eqref{slaxk1}-\eqref{slaxk2} do not form proper submanifolds
with respect to the cubic Poisson tensor \eqref{pot3}
\begin{align}
\notag \theta_3(L)dH &= \pobr{(L\star dH\star L)_{\me -1+k},L}_\star- L\star (\pobr{dH,L}_\star)_{\me 2-k}\star L\\
&= -\pobr{(L\star dH\star L)_{< -1+k},L}_\star+ L\star (\pobr{dH,L}_\star)_{<
2-k}\star L\qquad k=1,2 .
\end{align}
Nevertheless, the Dirac reduction can be applied. Again, the number of constraints
depends on $N$, so the reduction has to be considered separately for each $N$.

Let us now consider more precisely the transformations from the evolution systems
constructed from the algebra $\alga$ to the systems constructed from $A_\alpha$ for $
r=0$ and $r=1$. The linear transformation $D^\alpha:A_0 \longrightarrow A_\alpha$ is
simply given by \eqref{lintrans} as $\alga \cong A_0$. First consider the case of $r=0$. Let
\begin{equation}
\begin{split}
 L &= \sum_{m\me 0} u_m p^m +\sum_{m<0} p^m \star^0 u_m \in A_0\\
 L_\alpha &= \sum_{n\me 0} w_n p^n +\sum_{n<0} p^n \star^\alpha w_n \in A_{\alpha} .
\end{split}
\end{equation}
Then, $L_\alpha = D^\alpha L$, where $D^\alpha = e^{-\frac{\alpha}{2} \hk \pr_p \pr_x}$. As follows the dynamical fields are interrelated in the following way for $n\me 0$
\begin{equation}\label{ta}
\begin{split}
  w_n &= \sum_{s\geq 0} \bra{-\frac{\alpha}{2} \hk}^s \binom{n+s}{s} \bra{u_{n+s}}_{sx}\\
  u_n &= \sum_{s\geq 0} \bra{\frac{\alpha}{2} \hk}^s \binom{n+s}{s} \bra{w_{n+s}}_{sx} 
\end{split}
\end{equation}
and for $n<0$ $u_n = w_n$. We denote this transformation in the operator form by $w = \phi(u)$, then the
Fr\'{e}chet derivative of $\phi$, such that $\bra{w_m}_t = \sum_n \bra{\phi'}_m^n
\bra{u_n}_t$, is
\begin{equation}
\bra{\phi'}_m^n = \sum_{k\geq 0} \frac{\pr w_m}{\pr (u_n)_{kx}}\pr_x^k =
   \begin{cases}
   \bra{-\frac{\alpha}{2} \hk}^{n-m} \binom{n}{n-m}\pr_x^{n-m}  & \text{for $n\me m\me 0$}\\
   \delta_{m,n}                             & \text{for $m<0$}\\
      0                       & \text{for the rest}
   \end{cases}
\end{equation}
and its adjoint is
\begin{equation}
\bra{{\phi'}^\dagger}_n^m = \sum_{k\geq 0} (-1)^k \pr_x^k \frac{\pr w_m}{\pr (u_n)_{kx}} =
   \begin{cases}
   \bra{\frac{\alpha}{2} \hk}^{n-m} \binom{n}{n-m}\pr_x^{n-m}  & \text{for $n\me m\me 0$}\\
    \delta_{m,n}                                & \text{for $m<0$}\\
      0                       & \text{for the rest}
   \end{cases} .
\end{equation} Consider now the case of $r=1$. Let
\begin{equation}
 L = \sum_{m} u_m p^m\in A_0\qquad L_\alpha = \sum_{n} w_n p^n \in A_{\alpha} .
\end{equation}
Then, $L_\alpha = D^\alpha L$, where $D^\alpha = e^{-\frac{\alpha}{2} \hk p\pr_p \pr_x}$ and from
\eqref{mutrel} the relations between the dynamical fields are
\begin{equation}\label{tb}
\begin{split}
  w_n(x) &= \sum_{s\geq 0} \bra{-\frac{\alpha}{2} \hk}^s \frac{1}{s!} n^s \bra{u_n(x)}_{sx} = \e^{-n \frac{\alpha}{2}} u_n(x) = u_n\bra{x-n\frac{\alpha}{2} \hk}\\
  u_n(x) &= \sum_{s\geq 0} \bra{\frac{\alpha}{2} \hk}^s \frac{1}{s!} n^s \bra{w_n(x)}_{sx} = \e^{-n \frac{\alpha}{2}}w_n(x)= w_n\bra{x+n\frac{\alpha}{2} \hk} .
\end{split}
\end{equation}
Again, if we denote the transformation as $w = \phi(u)$, then
\begin{equation}
\bra{\phi'}_m^n = \delta_{m,n}\ \e^{-m \frac{\alpha}{2}}\quad \text{and} \quad
\bra{{\phi'}^\dagger}_m^n = \delta_{m,n}\ \e^{m \frac{\alpha}{2}} .
\end{equation}
Thus, obviously in both cases, when
\begin{equation}
 u_t = \theta dH\qquad w_t = \widetilde{\theta} d\widetilde{H}
\end{equation}
then
\begin{equation}\label{fl}
 w_t = \phi' u_t\qquad \widetilde{\theta} = \phi \theta {\phi'}^\dagger\qquad dH = {\phi'}^\dagger d\widetilde{H} .
\end{equation}

We will now display examples of some field and lattice soliton systems calculated in
the quantization scheme considered. We consider the Lax hierarchy \eqref{laxh} with
little changed numerations of evolution variables
\begin{equation}\label{hhh}
L_{t_n} = \pobr{(L^{\frac{n}{N}})_{\me -r+k},L}_\star
\end{equation}
where $N$ is the highest order of the Lax operator $L$. We will exhibit the first
nontrivial equation of the Lax hierarchy \eqref{hhh}. For simplicity we present only
the bi-Hamiltonian structure. The advantage of the use of $\alga$ algebra is that
during whole calculations there is no need of using the $\star^\alpha$-product in
explicit form and we only use the commutation relations \eqref{rule1}, \eqref{rule2}.
As a result, one gets the same equations and Poisson structures as these obtained from
quantized algebra $A_0$. The Hamiltonian systems related to quantized algebras
$A_\alpha$ are simply reconstructed via the linear transformation \eqref{ta},
\eqref{tb} and formulas \eqref{fl}. Such a procedure of calculations is applied in
present examples and we have written down only the final results for $A_\alpha$.

\begin{example}
The Boussinesq system: $r=0,k=0$.
\rm The dispersionless Boussinesq Hamiltonian systems is given in the form
\begin{equation}\label{dBouss}
\Matrix{t_2}{u}{v} = \Matrix{}{2v_x}{-\frac{2}{3}uu_x} =
\theta_{1}dH_1=\theta_2^{red}dH_2
\end{equation}
where the Poisson tensors are
\begin{equation}
\theta_{1} = 3 \pmatrix{cc}{0 & \Dx{}\\ \Dx{} & 0}\qquad \theta_2^{red} =
\pmatrix{cc}{\Dx{}u+u\Dx{} & 2\Dx{}v+v\Dx{}\\
\Dx{}v+2v\Dx{} & -\frac{2}{3}u\Dx{}u}
\end{equation}
and
\begin{equation} H_1 = \frac{1}{3} \int_\Omega (-\frac{1}{9}u^3+v^2)\ dx\qquad H_2
= \int_\Omega v\ dx .
\end{equation}
The system \eqref{dBouss} has the following Lax representation \cite{BS}
\begin{equation}
L_{t_2} = \pobr{\bra{L^{\frac{2}{3}}}_{\me 0},L}_{PB}^0
\end{equation}
for Lax operator in the form
\begin{equation}
L= p^3 + up + v.
\end{equation}
The quantization procedure leads now to the following Lax operator in $\alga$
\begin{equation}
L = p^3+ u\star p+ v.
\end{equation}
Then, one can derive the Boussinesq system from
\begin{eqnarray}
L_{t_2} = \pobr{\bra{L^{\frac{2}{3}}}_{\me 0},L}_\star.
\end{eqnarray}
Now, by the transformation to the algebras $A_\alpha$
one finds the following systems
\begin{align}\label{Bouss}
\notag \Matrix{t_2}{u}{v} &=  \Matrix{}{  2 v_x+\bra{\alpha -1}\hk u_{2x}   }{ -\frac{2}{3}u u_x +\bra{1-\alpha}\hk v_{2x}-\bra{\frac{\alpha^2}{2}-\alpha+\frac{2}{3}}\hk^2 u_{3x} }\\
&= \theta_1 dH_1 = \theta_2  dH_2 .
\end{align}
The respective Poisson tensors can be calculated from \eqref{p1r0k}
\begin{equation}
\theta_1 =  3 \pmatrix{cc}{
   0      &   \pr_x    \\
   \pr_x      &  0
 }\\
\end{equation}
and from \eqref{p2r0k0}
\begin{equation}
\theta_2^{red} = \frac{1}{2} \pmatrix{cc}{
\theta_{uu}      & \theta_{uv}       \\
-\bra{\theta^{uv}}^\dagger    &  \theta_{vv} }
\end{equation}
where
\begin{align}
\notag \theta_{uu} &= \pr_x u + u\pr_x+2\hk^2 \pr_x^3\\
\notag  \theta_{uv} &= 2\pr_x v +v\pr_x +\hk \brac{\alpha \pr_x u_x +\alpha u_x \pr_x
-\pr_x^2 u +\alpha u
\pr_x^2}+\bra{\alpha -1}\hk^3 \pr_x^4\\
\notag \theta_{vv} &= -\frac{2}{3}u\pr_x u +\bra{1-\alpha}\hk \brac{\pr_x^2 v -
v\pr_x^2}- \bra{\frac{\alpha^2}{4}-\frac{\alpha}{2}+\frac{2}{3}} \hk^2 \brac{\pr_x^3 u
+ u
\pr_x^3}\\
\notag &\quad - \frac{\alpha}{2} \bra{\frac{\alpha}{2}-1}\hk^2 \brac{\pr_x^2 u_x - u_x
\pr_x^2}-\bra{\frac{\alpha^2}{2}-\alpha+\frac{2}{3}}  \hk^4 \pr_x^5 .
\end{align}
The Hamiltonians are given in the following form
\begin{align}
H_1 &= \frac{1}{3} \int_\Omega \brac{ -\frac{1}{9}u^3+v^2+\bra{\alpha -1}\hk u_x v +
\frac{\alpha^2}{4} \hk^2 u_x^2 +\bra{\frac{\alpha}{2}-\frac{1}{3}}\hk^2 u u_{2x} }\ dx\\
H_2 &= \int_\Omega v\ dx.
\end{align}
In the case of $\alpha = 0$ \eqref{Bouss} is obviously the standard case of Boussinesq
system obtained from Gel'fand-Dikii hierarchy, for $\alpha=1$ it is the Moyal case.
The limit $\hk \rightarrow 0$ of \eqref{Bouss} gives \eqref{dBouss} as it should be.
\end{example}

\begin{example}
The Kaup-Broer (KB) system: $r=0,k=1$.

\rm The dispersionless system given by
\begin{equation}\label{benney}
\Matrix{t_2}{u}{v} = 2\Matrix{}{uu_x+v_x}{u_xv+uv_x} = \theta_{1}dH_1 =
\theta_2^{red}dH_2,
\end{equation}
where
\begin{equation}
\begin{split}
&\theta_{1} = \pmatrix{cc}{
0& \Dx{}\\
\Dx{}& 0\\
}\qquad \theta_2^{red} = \pmatrix{cc}{
2\Dx{}& \Dx{}u\\
u\Dx{}& \Dx{}v+v\Dx{}
}\\
&H_1 = \int_\Omega (u^2v+v^2)\ dx\qquad H_2 = \int_\Omega uv\ dx
\end{split}
\end{equation}
is known as the Benney system. The Lax representation for
\eqref{benney} is \cite{BS}
\begin{eqnarray}
L_{t_2} = \pobr{\bra{L^2}_{\me 0},L}_{PB}^0
\end{eqnarray}
where
\begin{equation}
L = p+ u + v p^{-1}.
\end{equation}
The quantized Lax operator in $\alga$ is
\begin{equation}
L = p+ u + p^{-1}\star v.
\end{equation}
We derive the KB system, which is the dispersive Benney system, from
\begin{eqnarray}
L_{t_2} = \pobr{\bra{L^2}_{\me 0},L}_\star.
\end{eqnarray}
And, by the transformation to the algebras $A_\alpha$
one gets the following systems
\begin{align}\label{KB}
\notag \Matrix{t_2}{u}{v} &= \Matrix{}{  2 u u_x+\bra{\alpha+1}\hk u_{2x}+2 v_x   }{   2\bra{u v}_x-\alpha \bra{1-\frac{\alpha}{2}}\hk^2u_{3x}-\bra{\alpha+1}\hk v_{2x} }\\
&= \theta_1 dH_1 = \theta_2  dH_2 .
\end{align}
The Poisson tensors are
\begin{equation}
\theta_1 =  \pmatrix{cc}{
   0      &   \pr_x    \\
   \pr_x      &  0
 }
\end{equation}
and
\begin{equation}
\theta_2 = \frac{1}{2}
\pmatrix{cc}{
\pr_x      & \pr_x u +\bra{\alpha+1}\hk\pr_x^2       \\
 u\pr_x-\bra{\alpha+1}\pr_x^2     &  v\pr_x+\pr_x v+\frac{1}{2}\alpha \hk \bra{u\pr_x^2-\pr_x^2 u}-\alpha \bra{1+\frac{\alpha}{2}}\hk^2 \pr_x^3
}.
\end{equation}
The Hamiltonians are
\begin{align}
H_1 &= \int_\Omega \brac{ u^2 v+v^2-\bra{\alpha+1}u v_x+\frac{\alpha}{2} \hk u^2 u_x+\frac{\alpha^2}{4} \hk^2 u_x^2 }\ dx\\
H_2 &= \int_\Omega \brac{ u v+\frac{\alpha}{2} \hk u u_x  }\ dx.
\end{align}
For $\alpha = 0$ \eqref{KB} is the standard case of KB system and
for $\alpha=1$ it is the Moyal case.
\end{example}

\begin{example}
Toda system: $r=1,k=1$.

\rm The dispersionless Toda system has the form
\begin{equation}\label{dToda}
\Matrix{t_1}{u}{v} = \Matrix{}{v_x}{u_xv} = \theta_{1}dH_1 = \theta_2^{red}dH_2,
\end{equation}
where
\begin{equation}
\begin{split}
&\theta_{1} = \pmatrix{cc}{
0& \Dx{}v\\
v\Dx{}& 0\\
}\qquad \pi_0^{red} = \pmatrix{cc}{
\Dx{}v+v\Dx{}& u\Dx{}v\\
v\Dx{}u& 2v\Dx{}v
}\\
&H_1 = \frac{1}{2} \int_\Omega (u^2+2v)\ dx\qquad H_2 = \int_\Omega u\ dx.
\end{split}
\end{equation}
The Lax representation for \eqref{dToda} is \cite{BS}
\begin{eqnarray}
L_{t_2} = \pobr{\bra{L^2}_{\me 0},L}_{PB}^1
\end{eqnarray}
where the Lax operator is
\begin{equation}
L = p+ u + v p^{-1}.
\end{equation}
Then, the quantization scheme leads to the following Lax operator in $\alga$ in the
form
\begin{equation}
L = p + u(x) + v(x)\star p^{-1}.
\end{equation}
One derive the Toda system from
\begin{eqnarray}
L_{t_2} = \pobr{\bra{L^2}_{\me 0},L}_\star.
\end{eqnarray}
Next, by the transformation to the algebras $A_\alpha$ 
one finds the following systems
\begin{align}\label{toda}
\notag \Matrix{t_2}{u(x)}{v(x)} &= \frac{1}{\hk} \Matrix{}{v\bra{x+\bra{1-\frac{\alpha}{2}}\hk}-v\bra{x-\frac{\alpha}{2}\hk}}{v(x)\brac{u\bra{x+\frac{\alpha}{2}\hk}-u\bra{x-\bra{1-\frac{\alpha}{2}}\hk}}}\\
&= \theta_1 dH_1 = \theta_2  dH_2 .
\end{align}
The respective Poisson tensors are
\begin{equation}
\theta_1 = \frac{1}{\hk}
\pmatrix{cc}{
0 & \brac{\e^{\bra{1-\frac{\alpha}{2}}}-\e^{-\frac{\alpha}{2}}}v(x)\\
v(x)\brac{\e^{\frac{\alpha}{2}}-\e^{-\bra{1-\frac{\alpha}{2}}}} & 0
}
\end{equation}
and
\begin{equation}
\theta_2^{red} = \frac{1}{\hk} \pmatrix{cc}{ \e^{ \bra{1-\frac{\alpha}{2}} } v(x) \e^{
\frac{\alpha}{2}} - \e^{-\frac{\alpha}{2}} v(x) \e^{ \bra{\frac{\alpha}{2}-1}} & u(x)
\brac{ \e^{ \bra{1-\frac{\alpha}{2}}} - \e^{-\frac{\alpha}{2}} } v(x)
\\
v(x) \brac{ \e^{ \frac{\alpha}{2}}-\e^{\bra{\frac{\alpha}{2}-1}} }u(x) & v(x) \brac{
\e^{}-\e^{-1} } v(x)
 }
\end{equation}
The Hamiltonians are
\begin{equation}
H_1 = \int_\Omega \brac{v\bra{x}+\frac{1}{2}u^2(x)}\ dx\qquad H_2 = \int_\Omega u(x)\
dx.
\end{equation}
The case of $\alpha = 0$ of \eqref{toda} is the standard case of Toda system, the case
of $\alpha=1$ is the Moyal case. Notice that in our construction Toda equation depends
on continuous coordinate $x$ contrary to a standard case when $x$ is integer.
\end{example}

\begin{example}
Three field system: $r=1,k=2$.

\rm The dispersionless system is given in the form
\begin{equation}\label{three}
\Matrixx{t_2}{u}{v}{w} = \Matrixx{}{2uw_x}{u_x+vw_x}{v_x} =
\theta_{1}dH_1=\theta_2^{red}dH_2
\end{equation}
where the Poisson tensors are
\begin{equation}
\begin{split}
\theta_{1} &= 3 \pmatrix{ccc}{ 0 & 0& 2u\pr_x \\ 0& \pr_x u +u\pr_x & v\pr_x\\
2\pr_x u & \pr_x v & 0}\\
\theta_2^{red} &= \pmatrix{ccc}{ 6u\pr_xu& 4u\pr_xv& 2u\pr_xw\\
4v\pr_xu& 2v\pr_xv+u\pr_xw+w\pr_xu& v\pr_xw+\pr_xu+2u\pr_x\\
2w\pr_xu& w\pr_xv+2\pr_xu+u\pr_x& \pr_xv+v\pr_x}
\end{split}
\end{equation}
and
\begin{equation} H_1 = \int_\Omega (v+\frac{1}{2} w^2)\ dx\qquad H_2
= \int_\Omega w\ dx .
\end{equation}
The system \eqref{three} has the following Lax representation
\begin{equation}
L_{t_2} = \pobr{\bra{L}_{\me 1},L}_{PB}^1
\end{equation}
for Lax operator in the form
\begin{equation}
L= u p^2 + vp + w+p^{-2}.
\end{equation}
The quantization procedure leads to the following Lax operator in $\alga$
\begin{equation}
L =  u(x)\star p^2 + v(x)\star p +w(x)+p^{-2}.
\end{equation}
Then, one derive the dispersive version of \eqref{three} from
\begin{eqnarray}
L_{t_2} = \pobr{\bra{L}_{\me 1},L}_\star
\end{eqnarray}
and by the transformation to the algebras $A_\alpha$
one finds the following systems
\begin{equation}
 \begin{split}
u(x)_{t_2} =& \frac{1}{\hk}  u(x)\brac{w\bra{x+\bra{2-\alpha}\hk}-w\bra{x-\alpha \hk}}      \\
v(x)_{t_2} =& \frac{1}{\hk}  \biggl [  u\bra{x+\frac{\alpha}{2} \hk} -u\bra{x+\bra{\frac{\alpha}{2}-1}\hk} \biggr .\\
& \qquad \biggl . +v(x)\brac{w\bra{x-\bra{\frac{\alpha}{2}-1}\hk}-w\bra{x-\frac{\alpha}{2} \hk}}  \biggr ]\\
w(x)_{t_2} =& \frac{1}{\hk} \brac{ v\bra{x+\frac{\alpha}{2} \hk} -v\bra{x+\bra{\frac{\alpha}{2}-1}\hk}    }.
 \end{split}
\end{equation}
The linear Poisson tensor is
\begin{equation}
\theta_1 = \frac{1}{\hk} \pmatrix{ccc}{
0 & 0 & \theta_1^{uw} \\
0 &  \theta_1^{vv}   &  \theta_1^{vw}  \\
-\bra{\theta_1^{uw}}^\dagger & -\bra{\theta_1^{vw}}^\dagger & 0
}
\end{equation}
where
\begin{align}\notag
\theta_1^{uw} & = u(x) \brac{\e^{\bra{2-\alpha}}-  \e^{-\alpha  } }\\\notag
\theta_1^{vv} & = \e^{\frac{\alpha}{2} } u(x) \e^{\bra{1-\frac{\alpha}{2}}} - \e^{\bra{\frac{\alpha}{2}-1}} u(x) \e^{-\frac{\alpha}{2} }\\\notag
\theta_1^{vw} & = v(x) \brac{ \e^{\bra{1-\frac{\alpha}{2}} } - \e^{-\frac{\alpha}{2}  } }.
\end{align}
The quadratic Poisson tensor is
\begin{equation}
\theta_2^{red}  = \frac{1}{\hk} \pmatrix{ccc}{
\theta_2^{uu} & \theta_2^{uv}  & \theta_2^{uw}\\
-\bra{\theta_2^{uv}}^\dagger & \theta_2^{vv} & \theta_2^{vw}\\
-\bra{\theta_2^{uw}}^\dagger  & -\bra{\theta_2^{vw}}^\dagger  & \theta_2^{ww}
 }
\end{equation}
where
\begin{align}\notag
\theta_2^{uu}  =& u(x)\brac{\e^{2}+\e-\e^{-1}-\e^{-2}} u(x)\\\notag \theta_2^{uv} =&
u(x)\brac{
\e^{\bra{2-\frac{\alpha}{2}}}+\e^{\bra{1-\frac{\alpha}{2}}}-\e^{-\frac{\alpha}{2}}-\e^{-\bra{1+\frac{\alpha}{2}}}
}v(x)\\\notag \theta_2^{uw} =& u(x)\brac{ \e^{\bra{2-\alpha} }-\e^{-\alpha }}
w(x)\\\notag \theta_2^{vv}  =& v(x)\brac{\e-\e^{-1}} v(x)+\e^{\frac{\alpha}{2}} u(x)
\e^{\bra{1-\alpha}} w(x)\e^{\frac{\alpha}{2}}\\\notag &-\e^{-\frac{\alpha}{2}} w(x)
\e^{\bra{\alpha -1}} u(x) \e^{-\frac{\alpha}{2}}\\\notag \theta_2^{vw} =& v(x)\brac{
\e^{\bra{1-\frac{\alpha}{2}}}-\e^{-\frac{\alpha}{2}}}w(x)+\e^{\frac{\alpha}{2}} u(x)
\e^{\bra{2-\alpha}} - \e^{\bra{\frac{\alpha}{2} -1}} u(x) \e^{-\alpha }\\\notag
\theta_2^{ww}  =& \e^{\frac{\alpha}{2}} v(x) \e^{\bra{1-\frac{\alpha}{2}}} -
\e^{\bra{\frac{\alpha}{2} -1}} v(x) \e^{ -\frac{\alpha}{2}} .
\end{align}
The respective Hamiltonians have the form
\begin{equation}
H_1 =  \int_\Omega \brac{v(x) +\frac{1}{2} w(x)^2    }\ dx\qquad H_2 = \int_\Omega
w(x)\ dx.
\end{equation}
The case with $\alpha=0$ ($\hk=1$) and integer $x$ was constructed in \cite{BM}.

\end{example}

\section{Conclusions}

In this paper we have presented a systematic construction of the field and lattice
soliton systems from underlying multi-Hamiltonian dispersionless systems. Actually,
the passage has been made on the level of appropriate Lax representations through the
Weyl-Moyal like deformation quantization procedure. In a previous paper \cite{BS} we
have constructed Lax representations for a wide class of dispersionless systems with
multi-Hamiltonian structures, derived from classical $R$-matrix theory. The number of
the constructed dispersionless systems is much greater then the number of known soliton
systems (dispersive integrable systems). So, the question rises whether for any
dispersionless Lax hierarchy one can construct a related soliton hierarchy. We have
tried to obtain an answer to this question via the procedure of deformation
quantization for Poisson algebras of dispersionless systems and appropriate $R$-matrix
theory. We have managed to quantize all Poisson algebras (with arbitrary $r$
\eqref{por}) but the $R$-matrix, at least of the form \eqref{rp}, exists only in the
case of two algebras, i.e. for $r=0$ and $r=1$, respectively. The first case leads to
soliton field systems related to the algebra of pseudo-differential operators
\eqref{pdo}, and the second one leads to lattice soliton systems related to the
algebra of shift operators \eqref{so}. In that sense, although we have failed to
construct new soliton equations through presented deformation procedure, nevertheless
we have found a unified procedure of the construction of field and lattice Hamiltonian
soliton systems in one scheme.


\begin{thebibliography}{99}

\bibitem{Ta} Takasaki K 1994 Non-abelian KP hierarchy with Moyal algebraic coefficients {\it J. Geom. Phys.} {\bf 14} 322-64

\bibitem{S1} Strachan I A B 1995 The Moyal bracket and the dispersionless limit of the KP hierarchy {\it J. Phys. A.} {\bf 20} 1967-75

\bibitem{S2} Strachan I A B 1997 A geometry for multidimensional integrable systems {\it J. Geom. Phys.} {\bf 21} 255-278

\bibitem{DP} Das A and Popowicz Z 2001 Properties of Moyal-Lax representation {\it Phys. Lett.} B {\bf 510} 264-70

\bibitem{TLC} Tu M-H, Lee N-C and Chen Y-T 2002 Conformal covariantization of Moyal-Lax operators {\it J. Phys A: Math. Gen.,} {\bf 35} 4375-84

\bibitem{S-T-S} Semenov-Tian-Shansky M A 1983 What is a classical r-matrix?  {\it Funct. Anal. Appl.} {\bf 17} 259

\bibitem{OR} Oevel W and Ragnisco O 1990 R-matrices and higher Poisson brackets for integrable systems {\it Physica} {\bf 161 A} 181

\bibitem{KO} Konopelchenko B G and Oevel W 1993 An r-matrix approach to nonstandard classes of integrable equations {\it Publ. RIMS, Kyoto Univ.} {\bf 29} 581-666

\bibitem{O2} Oevel W. and Strampp W. 1993 Constrained KP hierarchy and bi-Hamiltonian structures {\it Commun. Math. Phys.} {\bf 157} 51

\bibitem{O3} Oevel W 1994 A note on the Poisson brackets associated with Lax operators {\it Phys. Lett.} A {\bf 186} 79

\bibitem{Bl1} B\l aszak M 1998 {\it Multi-Hamiltonian Theory of Dynamical Systems} (Berlin: Springer)


\bibitem{K} Kupershmidt B A 1985 Discrete Lax equations and differential-difference
calculus {\it Asterisque} {\bf 123}

\bibitem{Su} Suris Y B 1993 On the bi-Hamiltonian structure of Toda and relativistic Toda lattices {\it Phys. Lett.} A {\bf 180} 419

\bibitem{BM} B\l aszak M and Marciniak K 1994 R-matrix approach to lattice integrable
systems {\it J. Math. Phys.} {\bf 35} 4661

\bibitem{O} Oevel W 1996 Poisson Brackets in Integrable Lattice Systems in {\it
Algebraic Aspects of Integrable Systems} edited by A S Fokas and I M Gelfand {\it
Progress in Nonlinear Differential Equations} Vol. 26 (Birkh\"{a}user-Boston) 261

\bibitem{T-T} Takasaki K and Takebe T 1995 Integrable hierarchies and dispersionless limit {\it Rev. Math. Phys.} {\bf 7} 743-808

\bibitem{Li} Li Luen-Chau 1999 Classical r-Matrices and Compatible Poisson Structures for Lax Equations in Poisson Algebras {\it Commun. Math. Phys.} {\bf 203} 573-92

\bibitem{Bl2} B\l aszak M 2002 Classical R-matrices on Poisson algebras and related dispersionless systems {\it Phys. Lett.} A {\bf 297} 191-5

\bibitem{BS} B\l aszak M and Szablikowski B M 2002 Classical $R$-matrix theory of dispersionless systems: I. (1+1)-dimension theory {\it J. Phys A: Math. Gen.,} {\bf 35} 10325-44

\bibitem{BFFLS} Bayen F, Flato M, Fronsdal C, Lichnerowicz A and Sternheimer D 1978 Deformation theory and quantization. I. Deformations of symplectic structures {\it Ann. Physics.} {\bf 111} 61-110

\bibitem{WL} De Wilde M and Lecomte P B A 1983 Existence of star-products and of formal deformations in Poisson Lie algebra of arbitrary symplectic manifolds {\it Lett. Math. Phys.} {\bf 7} 487-96

\bibitem{Fe} Fedosov B 1994 A simple geometric construction of deformation quantization {\it J. Diff. Geom.} {\it 40} 213-38

\bibitem{Kon} Kontsevich M 1997 Deformation quantization of Poisson manifolds. I. {\it Preprint} q-alg/9709040

\bibitem{G} Groenewold H 1946 On the principles of elementary quantum mechanics {\it Physica} {\bf 12} 405

\bibitem{M} Moyal J 1949 Quantum mechanics as a statistical theory {\it Proc. Camb.
Phil. Soc.} {\bf 45} 99-124

\bibitem{Ma} Manin Yu I 1979 {\it J. Sov. Math.} {\bf 11} 1-122

\bibitem{K2} Kupershmidt B A 1990 Quantizations and integrable systems {\it Lett. Math.
Phys.} {\bf 20} 19-31




\end{thebibliography}
\end{document}